Author: **Dilip  G. Banhatti**   302102          Affiliation: Madurai-Kamaraj University, India

Email: dilip.g.banhatti@gmail.com, banhatti@uni-muenster.de

Abstract: The example of disk galaxy rotation curves is given for inferring dark matter from redundant computational procedure because proper care of astrophysical and computational context was not taken. At least three attempts that take the context into account have not found adequate voice because of haste in wrongly concluding existence of dark matter on the part of even experts. This firmly entrenched view, prevalent for about 3/4ths of a century, has now become difficult to correct. Context-awareness must be borne in mind at every step to avoid such a situation. Perhaps other examples exist.

---------------
Manuscript

**Astrophysical (and cosmological) context**

Gravitating matter too faint to detect with state-of-art photon-detecting astronomical instruments has evolved from being called *missing mass* in the 1930s **[1]** to *missing light* four decades later to *dark matter* soon afterwards. The concept was originally invented to ensure virial energy balance in clusters of galaxies. It was tested on smaller (galaxy groups' and single-galaxy) scales and larger (observable universe) scales in a gravity-research-foundation award-winning entry **[2]**, and later acquired a memetic life of its own in many astrophysical (and cosmological) contexts. It was soon joined by its sister concept *dark energy*.

**Computation on individual galaxy scale**

Galaxy formation scenarios and relevant observations indicate that interactions between galaxy-sized matter agglomerations may be an essential feature of the formation of individual galaxies. However, to a first approximation, individual galaxies (even if binaries or members of larger groups) may be taken to be isolated stable dynamical systems. The large size of an average individual galaxy warrants taking proper account of the different arrival times of photons starting their journey (quasi-)simultaneously from its different parts. That we can study whole galaxies as individual objects means that galaxian orbital and evolutionary timescales are much longer than the differences in light-travel times between photons originating from two sides of a galaxy. There is also the question of propagation of gravitational interaction between different parts of a galaxy. Newton's laws of gravitation and dynamics assume simultaneous inter-particle action, while general relativity properly applied should incorporate the local and global space-time structure relative to the relevant matter-energy tensor. As seen for Mercury's orbit in the inner solar system, general relativity theory is needed to tally with observations (within errors), Newtonian dynamics and gravitation even then being a very good approximation. It may, therefore, be applied, at least as a first approximation, on the galaxian scale. Applying Newtonian dynamics and gravitation has been the practical recourse of choice for most astrophysicists studying disk galaxy dynamics or any other topic in galaxian dynamics.

**Three attempts to get disk galaxy dynamics right**

**(1) Fourier numerical method in cylindrical polar coordinates [3]**

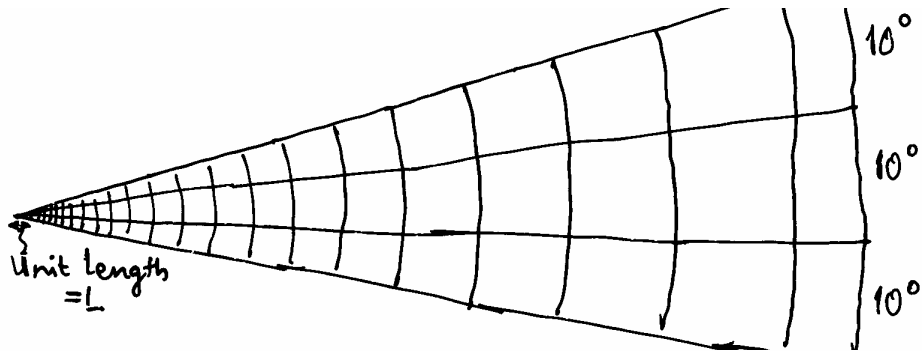

The disk plane is defined as z = 0 in a coordinate system with the radial coordinate logarithmic to allow for central mass concentration. Effects of disk thickness and of any halo must be taken into account by projection onto the base coordinate plane z = 0. Newtonian gravitation (that is, force proportional to inverse-squared distance) is used between any two of 60000 particles, with a smoothing parameter added to the interparticle distance to avoid infinity at coincidence of the two particles. Furthermore, leapfrog numerical scheme, also called Verlet method in molecular dynamics, is used for fast but accurate convergence. The actual operative calculations are carried out in the Fourier domain using fast Fourier transform to switch between the physical and Fourier domains. This overall numerical scheme has been used for various aspects of disk galaxies including spiral arms, tidal interactions and Seyfert activity. The scheme is quite efficient and needs to be used to estimate galaxian physical parameters from observed rotation curves without a priori assuming any dark matter in the halo or (especially) the disk.

**(2) Matrix inversion method in cylindrical polar coordinates [4]**
The disk plane is designated z = 0 with 250000 particles distributed along 500 rings of radii proportional to $i^2$, where i is the ring number from centre outward, upto a finite outer radius $R_g$ of the galaxy disk. Each ring is assumed infinitely thin in the z-direction. Newton's force equations are first written in finite element form, and then matrix-inverted to get the masses $m_i$ of different rings, i = 0, 1, 2, ..., n; along with the constraint equation giving the sum over $m_i$ to be the total mass $M_g$. The method is tested with known analytic solutions before applying it to astronomical data for Milky Way Galaxy. This numerical scheme is also eminently suitable for wider use.

**(3) Mass distribution to rotation curve and vice versa (refs quoted by [5])**
The forward problem of finding the rotation curve from the mass distribution of a thick disk of a given profile as well as the reverse one of finding the mass distribution from a given rotation curve is solved, judiciously using finite-element techniques on a cylindrical polar coordinate grid. The number of axisymmetric rings used can be as few as 5, and needs to be upto about 100 for good results. The numerical scheme is tested on known analytic results, and then applied to a few disk galaxies for which astronomical data are readily available.

## Discussion

The first scheme was not really used for rotation curve ↔ mass distribution calculations, but is very suitable for such a purpose. The second and third were explicitly designed for the purpose, but the authors could not get a hearing from mainstream practitioners, who had in the meantime amassed an enormous volume of publications based on redundant computational procedure, as especially brought out by the author of the third scheme. In retrospect, it seems that mainstream researchers in this area are not sufficiently context-aware to retract their error. For a broader discussion of disk galaxy rotation curves vis-à-vis dark matter distribution, see, for example, ref. **[6]**.

## Acknowledgements

Gautam, Ranjani & Radha Banhatti's comments improved the presentation.

------------------------end of manuscript-----------------------------